\documentclass{svjour3}                     
\smartqed  
\usepackage{graphicx}
\begin{document}

\title{ Recent topics of mesic atoms and mesic nuclei\\
--  $\phi$ mesic nuclei exist ?--}


\author{Junko Yamagata-Sekihara         \and \\
Satoru Hirenzaki \and
Daniel Cabrera \and \\ Manuel J. Vicente Vacas
}


\institute{J. Yamagata-Sekihara \at
              Department of Physics, Nara Women's University, Nara 630-8506, Japan \\
              Tel.: +81-742-203373\\
              Fax: +81-742-203373\\
              \email{aaj.yamagata@cc.nara-wu.ac.jp}           
           \and
           S. Hirenzaki \at
              Department of Physics, Nara Women's University
\and
D. Cabrera \at
Departamento de F\'{\i}sica Te\'orica II,
Universidad Complutense, Madrid
\and
M. J. Vicente Vacas \at
Departamento de F\'{\i}sica Te\'orica and IFIC,
Centro Mixto Universidad de Valencia-CSIC
}

\date{Received: date / Accepted: date}

\maketitle

\begin{abstract}
We study $\phi$-meson production in nuclei to investigate the in-medium modification of the $\phi$-meson spectral function at finite density.
We consider (${\bar p},\phi$), ($\gamma,p$) and ($\pi^-,n$) reactions to produce a $\phi$-meson inside the nucleus and evaluate the effects of the medium modifications to reaction cross sections.
The structures of the bound states, $\phi$-mesic nuclei, are also studied.
For strong absorptive interaction cases, we need to know the spectrum shape in a wide energy region to deduce the properties of $\phi$.
\end{abstract}

\section{Introduction}
\label{sec:1}
Mesic atoms such as pionic- and kaonic-atoms are Coulomb assisted meson-nucleus bound systems and have been studied systematically for a long time~\cite{ref:1}. The binding energies and widths of these bound states provide us unique and valuable information on the meson-nucleus interactions.  Mesic nuclei, another kind of meson-nucleus systems, are meson-nucleus bound states mainly due to attractive strong interactions and have also been studied intensively in these days.  

In the contemporary hadron-nuclear physics, mesic atoms and mesic nuclei are considered to be very interesting objects in the following two aspects. First, they are {\it strongly interacting exotic many body systems} which should be explored by nuclear physicists. 
Second, mesic atoms and mesic nuclei provide unique laboratories for the studies of {\it hadron properties at finite density} which are significantly important to explore various aspects of the symmetries of the strong interaction~\cite{ref:2,ref:3}. 

In this letter, we will report our studies on structure and formation of $\phi$ mesic nuclei~\cite{ref:4}.  
In-medium properties of the $\phi$ meson have been studied theoretically, and they have a close relation to $K$ and $\bar{K}$ meson properties in medium because of the strong $\phi \rightarrow  K\bar{K}$ coupling.  The study of QCD sum rules~\cite{ref:5} and the data taken at KEK~\cite{ref:6} suggested a 3
$\%$ mass reduction of $\phi$ at the normal nuclear density, while the $\phi$ meson selfenergy calculated in Refs.~\cite{ref:7,ref:8} indicated a significantly smaller attractive potential for $\phi$. 
Since we can expect to obtain new information which is complementary to the invariant mass measurements~\cite{ref:6}, we investigate the $\phi$ mesic nuclear states to study the $\phi$ properties in medium. We will show the calculated results of the bound states and formation spectra for some reactions in the following sections which will help to consider experimental feasibilities.

\section{Optical potential and bound states}
\label{sec:2}
\begin{figure}
  \includegraphics[width=1\textwidth]{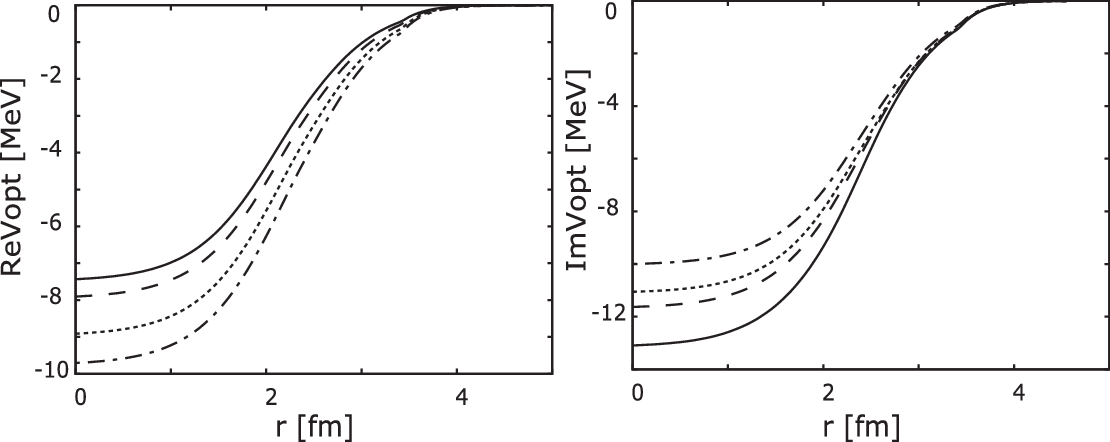}
\caption{ The $\phi$ meson optical potentials as a function of the radial coordinate $r$ for $\phi$ -$^{11}$B systems obtained from $\Pi_\phi$ in Ref.~\cite{ref:8}.
The left and right panels show the real and imaginary parts.
The solid, dashed, dotted, dotted-dashed lines indicate the potential strength for the $\phi$ meson energies Re $E-m_\phi=0$ MeV, $-10$ MeV, $-20$ MeV and $-30$ MeV, respectively.}
\label{fig:Vopt}       
\end{figure}

We show the $\phi$ meson optical potential $V_{\rm opt}(r)$ in Fig.~\ref{fig:Vopt}, which is obtained from the $\phi$ meson selfenergy $\Pi_\phi$ in Ref.~\cite{ref:8} as $V_{\rm opt}(r)=\displaystyle{\frac{1}{2\mu}}\Pi_\phi(\rho(r))$ within the local density approximation.
The selfenergy $\Pi_\phi$ depends on the energy of the $\phi$ meson and on the density $\rho$.

As shown in Fig.~\ref{fig:Vopt}, the theoretical potential based on $\Pi_\phi$~\cite{ref:8} is a weak attractive potential $V(0)\sim -7.5$ MeV at $E=m_\phi$.
To simulate the 3$\%$ mass reduction in nucleus indicated in Refs.~\cite{ref:5,ref:6}, we multiply a factor to the real part of the theoretical potential to scale Re $V(0)=-30$ MeV at $E=m_\phi$.
We use both theoretical (shallow) and scaled (deep) potentials in this article and compare the results to know the sensitivity of the observables to the potential depth.

We calculate the binding energies and widths of the bound states by solving the Klein-Gordon equation selfconsistently for the real part of the $\phi$ meson energy.
The calculated results are shown in Table~\ref{tab:1}.
\begin{table} 
\begin{center}
\caption{Calculated binding energies and widths of $\phi$ meson bound states in $^{11}$B,~$^{39}$K,~$^{123}$In,~and~$^{207}$Tl.
Widths do not include the $\phi$ decay width in vacuum.
}
\label{tab:1}       
\begin{tabular}{lll}
\hline\noalign{\smallskip}
 & Shallow potential & Deep potential \\
 & B.E.[MeV]($\Gamma$[MeV]) & B.E.[MeV]($\Gamma$[MeV]) \\
\noalign{\smallskip}\hline\noalign{\smallskip}
$^{11}$B & none & $1s$~~10.2(17.9) \\
$^{39}$K & none & $1s$~~28.5(20.0) \\
& & $2p$~~11.8(19.8) \\
$^{123}$In & $1s$~~2.34(21.6) & $1s$~~34.5(18.7) \\
& & $2s$~~12.2(19.9) \\
& & $2p$~~26.3(19.9) \\
$^{207}$Tl&$1s$~~3.73(22.5) & $1s$~~40.1(17.5) \\
& & $2s$~~24.4(20.2) \\
& & $2p$~~27.1(20.3) \\
& & $3p$~~13.1(20.4) \\
& & $3d$~~27.5(19.8) \\
\noalign{\smallskip}\hline
\end{tabular}
\end{center}
\end{table}
We see that there exist several bound states of $\phi$ in nuclei, however, the widths of the states are large and will make it difficult to observe any peak structure in the missing mass spectra.

\section{Formation spectra of $\phi$-nucleus systems}
\label{sec:3}
We use the Green's function method to calculate the formation spectra of the $\phi$ meson from nucleus~\cite{morimatsu85} as in the cases of other mesic nucleus formation~\cite{gata06,nagahiro05,jido02}.

\begin{figure}[htbp]
  \includegraphics[width=0.70\textwidth]{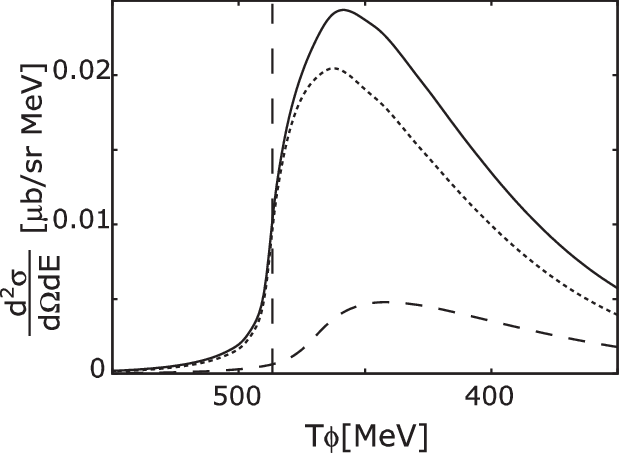}
\caption{Calculated spectra for the $\phi$-nucleus systems formation plotted as a function of the emitted $\phi$ meson energy in the $^{12}$C(${\bar p},\phi$) reaction at $p_{\bar p}=1.3$ GeV/c.
The dashed, dotted lines show the contributions from different proton hole states.
The vertical dashed line indicates the $\phi$ meson production threshold. 
The theoretical (shallow) optical potential is used.}
\label{fig:pbarp_chiral}       
\end{figure}

\begin{figure}[htbp]
  \includegraphics[width=0.70\textwidth]{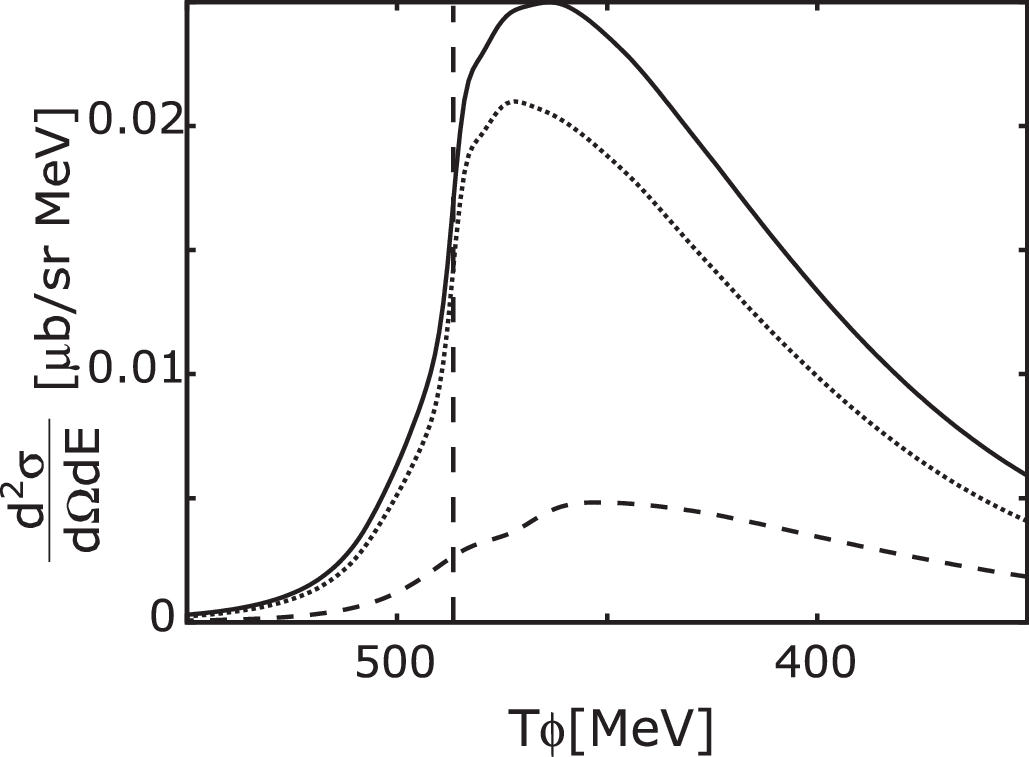}
\caption{ Calculated spectra for the $\phi$-nucleus systems formation plotted as a function of the emitted $\phi$ meson energy in the $^{12}$C(${\bar p},\phi$) reaction at $p_{\bar p}=1.3$ GeV/c.
The dashed, dotted lines show the contributions from different proton hole states.
The vertical dashed line indicates the $\phi$ meson production threshold. 
The scaled (deep) optical potential is used.}
\label{fig:pbarp_deep}       
\end{figure}

We show the calculated cross sections of the $^{12}$C(${\bar p},\phi$) reaction proposed in Ref.~\cite{iwasaki_pro} for the formation of a $\phi$ -$^{11}$B system in Fig.~\ref{fig:pbarp_chiral} for the theoretical (shallow) optical potential case.
The spectra shown in this article are folded spectra with the Gau$\beta$ian distribution with $\phi$ meson decay width in vacuum at $E=m_\phi$.
In this case, since no bound state exists as shown in Table~\ref{tab:1}, the spectrum has a smooth shape without any peak structures.
In Fig.~\ref{fig:pbarp_deep}, we show the same reaction spectra calculated with the scaled (deep) optical potential.
We find some enhancement of the spectra in the bound energy region for the deeper potential.
However, we do not observe any peak structures again for the deeper potential case even when there exists one bound state.
As we have expected from the widths of the bound states, it is difficult to observe any peak structure in the missing mass spectra.

\begin{figure}[htbp]
  \includegraphics[width=0.70\textwidth]{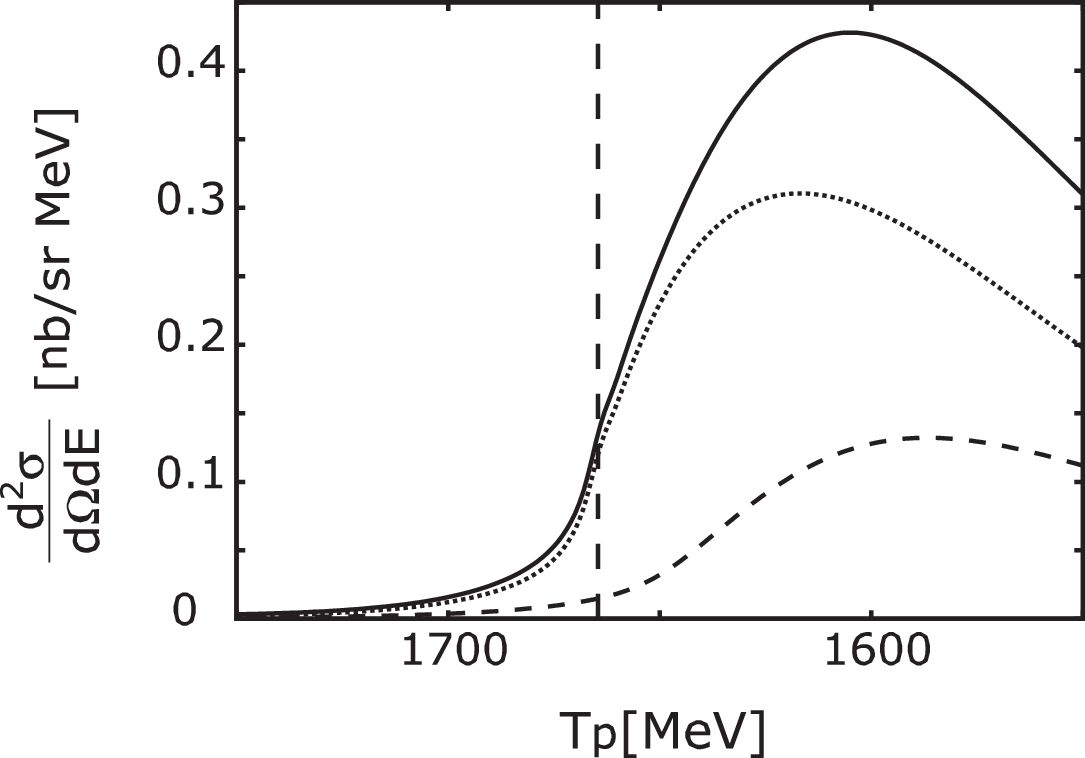}
\caption{Calculated spectra for the $\phi$-nucleus systems formation plotted as a function of the emitted proton energy in the $^{12}$C($\gamma,p$) reaction at $p_{\gamma}=2.7$ GeV/c.
The dashed, dotted lines show the contributions from different proton hole states.
The vertical dashed line indicates the $\phi$ meson production threshold.
The theoretical (shallow) optical potential is used. }
\label{fig:gamma_chiral}       
\end{figure}

\begin{figure}[htpb]
  \includegraphics[width=0.70\textwidth]{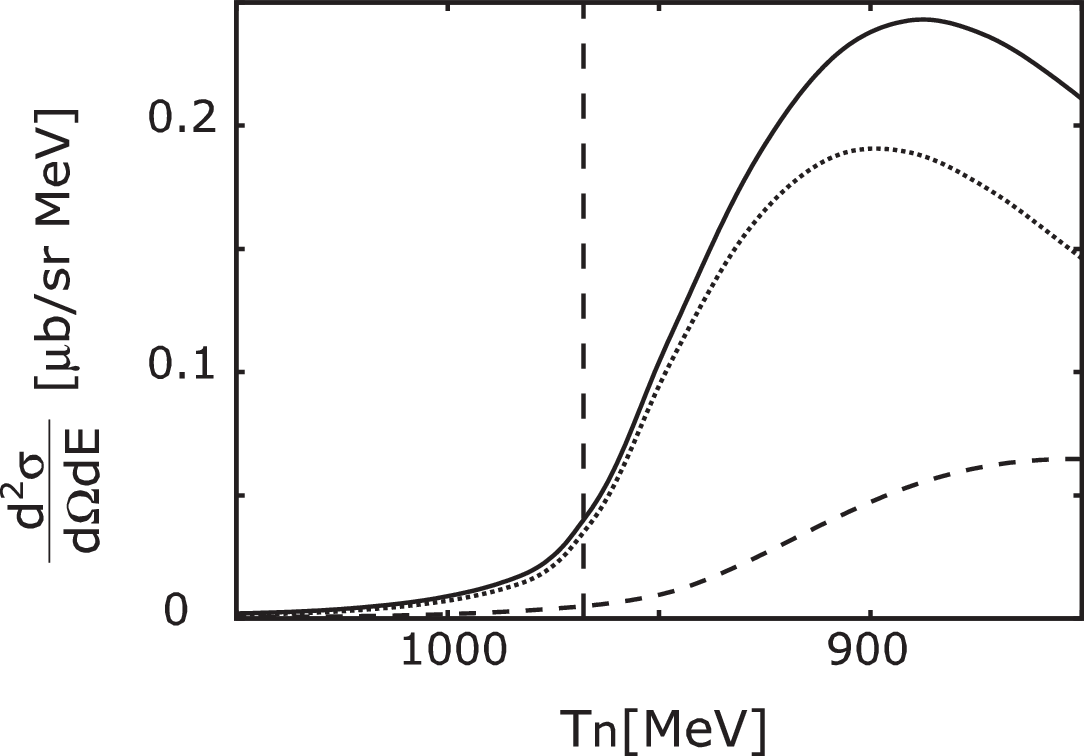}
\caption{Calculated spectra for the $\phi$-nucleus systems formation plotted as a function of the emitted neutron energy in the $^{12}$C($\pi^-,n$) reaction at $p_{\pi^-}=2.0$ GeV/c.
The dashed, dotted lines show the contributions from different proton hole states.
The vertical dashed line indicates the $\phi$ meson production threshold. 
The theoretical (shallow) optical potential is used.}
\label{fig:pin_chiral}       
\end{figure}

\begin{figure}[htpb]
  \includegraphics[width=0.70\textwidth]{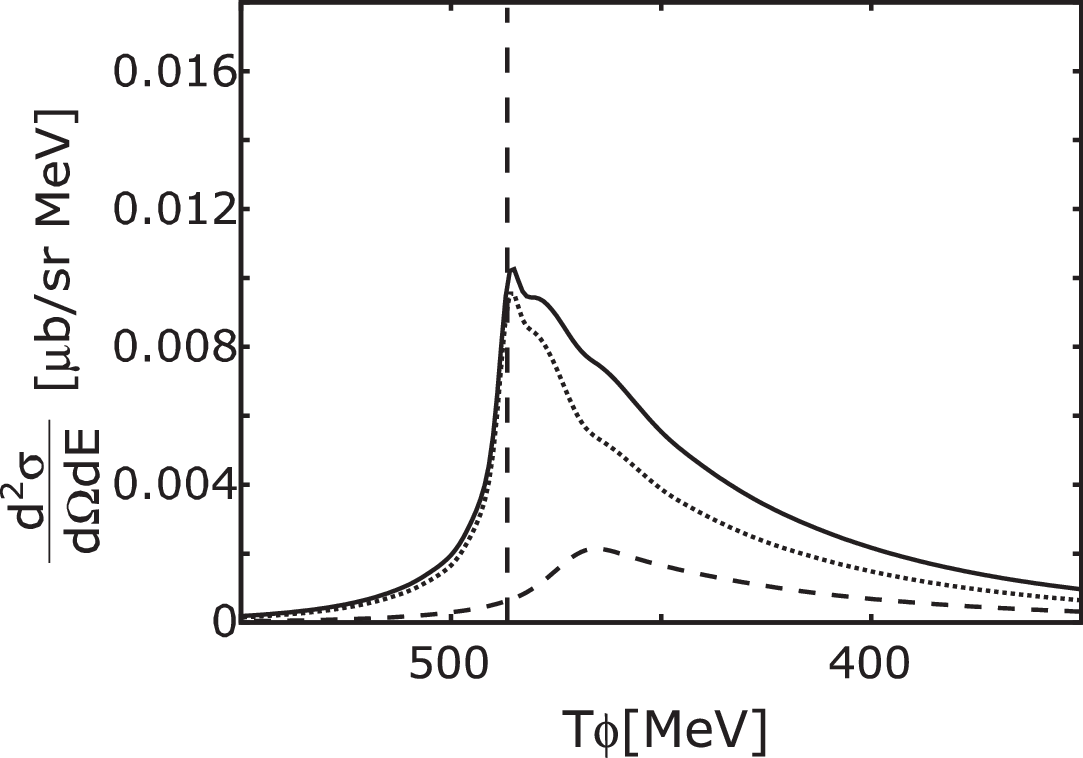}
\caption{The conversion part of the calculated spectra for the $\phi$-nucleus systems formation plotted as a function of the emitted $\phi$ meson energy in the $^{12}$C(${\bar p},\phi$) reaction at $p_{\bar p}=1.3$ GeV/c.
The dashed, dotted lines show the contributions from different proton hole states.
The vertical dashed line indicates the $\phi$ meson production threshold.
The theoretical (shallow) optical potential is used. }
\label{fig:pbarp_chiral_scon}       
\end{figure}

In Figs.~\ref{fig:gamma_chiral} and \ref{fig:pin_chiral}, we show the calculated results for different formation reactions, namely, $^{12}$C($\gamma,p$)~(Fig.~\ref{fig:gamma_chiral}) and $^{12}$C($\pi^-,n$) (Fig.~\ref{fig:pin_chiral}).
In both cases, the momentum transfers of the reactions are larger than for the ${\bar p}$ induced case and, hence, the spectra around threshold are suppressed in these reactions.
We also show the conversion part of the $^{12}$C(${\bar p},\phi$) spectra in Fig.~\ref{fig:pbarp_chiral_scon} which correspond to the (${\bar p},\phi$) spectra coincident with the particle emissions from $\phi$ meson decay (absorption) in nucleus.
One of the advantages of the (${\bar p},\phi$) reaction is a possible background reduction by the conversion spectrum as described in Ref.~\cite{iwasaki_pro}.
Thus, the conversion spectrum in Fig.~\ref{fig:pbarp_chiral_scon} will be an important piece of information to observe the $\phi$ meson properties in nucleus.
The details of our results shown in this article will be given in Ref.~\cite{ref:4}.

\section{Summary}
We study $\phi$-meson production in nuclei in this article and show the numerical results of $\phi$ mesic nuclear states and formation spectra for (${\bar p},\phi$), ($\gamma,p$) and ($\pi^-,n$) reactions.
Due to the large decay width of $\phi$ meson in nucleus, we have found that a clear peak observation in the missing mass spectra will be difficult.
Thus, we think that further theoretical investigations are very important to extract reliable information from experimental data.

\label{sec:4}




\end{document}